\documentclass{article}
\usepackage{amsfonts}
\usepackage{makeidx}
\usepackage{amssymb}
\usepackage{amsmath}
\usepackage{amstext}
\usepackage{graphicx}
\usepackage{latexsym}
\usepackage{amsthm}

\newtheorem{theorem}{Theorem}

\newtheorem{axiom}{Axiom}

\newtheorem{definition}[theorem]{Definition}

\newtheorem{proposition}[theorem]{Proposition}

\input tcilatex
\begin{document}

\title{Non-Archimedean Mathematics and the formalism of Quantum Mechanics}
\author{Vieri Benci \\
Mathematics, University of Pisa.}
\maketitle

\begin{abstract}
This paper is divided in four parts. In the introduction, we discuss the
program and the motivations of this paper. In section \ref{lt}, we introduce
the non-Archimedean field of Euclidean numbers $\mathbb{E}$ and we present a
summary of the theory of $\Lambda $-limits which can be considered as a
different approach to nonstandard methods. In the third part (section \ref%
{Ultrafunctions}), we define axiomatically the space of ultrafunctions which
are a kind of generalized function based on the field of Euclidean numbers $%
\mathbb{E}$. Finally, we describe an application of the previus theory to
the formalism of classical Quantum Mecanics.

\bigskip 

$\ \ \ \ \ \ \ \ \ \ \ \ \ \ \ \ $\ \ \ \ \ \ \ \ \ \ \ \ \ \ \ \ \ \ \ \ 
\textbf{Sommario}\newline

Questo articolo \`{e} diviso in quattro parti. Nell'introduzione discutiamo
il programma e le motivazioni di questo lavoro. Nella sezione \ref{lt},
introduciamo il campo non-Archimedeico dei numeri Euclidei $\mathbb{E}$ e
presentiamo un riassunto della teoria dei $\Lambda $-limiti che pu\`{o}
essere considerata come un approccio diverso ai metodi non standard. Nella
terza parte (sezione \ref{Ultrafunctions}), definiamo assiomaticamente lo
spazio delle ultrafunzioni che sono una sorta di funzione generalizzata
basate sul campo dei numeri euclidei $\mathbb{E}$. Infine, descriviamo
un'applicazione della della teoria precedente al formalismo della meccanica
quantistica classica.

\noindent \noindent \textbf{Parole chiave}. Ultrafunzioni, funzione Delta,
matematica non archimedea, analisi non standard, meccanica quantistica,
operatori autoaggiunti.
\end{abstract}

\tableofcontents

\section{Introduction}

This is not an article of Mathematics, but rather about Mathematics, or more
precisely about non-Archimedean Mathematics, namely the Mathematics based on
the use of infinitesimal and infinite numbers (NAM). We are convinced that
NAM deserves more attention than that which is generally attributed to it.
In this paper, we will expose some recent results on Euclidean numbers, $%
\Lambda $-theory and ultrafunctions emphasizing the ideas, but we have not
given the proofs of most results. In any case we have referred to the
original papers for the interested reader. Moreover, we will give an
original application to the formalism of Quantum Mechanics (QM).

\subsection{The Non-Archimedean Mathematics}

\begin{quotation}
\textit{La filosofia \`{e} scritta in questo grandissimo libro che
continuamente ci sta aperto innanzi a gli occhi (io dico l'universo), ma non
si pu\`{o} intendere se prima non s'impara a intender la lingua, e conoscer
i caratteri, ne'quali \`{e} scritto. Egli \`{e} scritto in lingua
matematica, e i caratteri son triangoli, cerchi, \textbf{spazi di Hilbert,
variet\`{a} simplettiche, frattali, infinitesimi ed altre entit\`{a}
matematiche }senza i quali mezzi \`{e} impossibile a intenderne umanamente
parola; senza questi \`{e} un aggirarsi vanamente per un oscuro laberinto.}
\end{quotation}

\qquad \textit{Galileo Galilei, Il Saggiatore (1623), Cap. VI}$\bigskip $%
\bigskip

We begin this article by recalling this passage by Galileo which highlights
one of the most fascinating aspects of mathematics: its ability to construct
models that allow us to understand, at least in part, the natural world.
Obviously, the words written in bold have been added to the original text in
order to emphasize the discovery of new mathematical entities that have
developed Galileo's ideas up to the current progress of science. Moreover,
these added words remind another aspect of the history of science (which we
do not fully understand): the ousting of infinitesimal quantities. The
scientific community has always accepted new mathematical entities,
especially if these are useful in the modeling of natural phenomena and in
solving the problems posed by the technique. Some of these entities are the
infinitesimals that, by the discovery of the infinitesimal calculus, have
been a carrier of the modern science. But despite the successes achieved
with their employment, they have been opposed and even fought by a
considerable part of the scientific community. We refer to the essay by Amir
Alexander \cite{amir} (see also \cite{bair2}) which tells how the Jesuits in
Italy and part of the \textit{Royal Society} in England fought the spread of
these "subversive notions". Sometimes it is said that people opposed
infinitesimals because of their lack of rigor, but this argument convinces
me little (see also \cite{Bla}). When at the end of the 19th century they
were placed on a more rigorous basis thanks to the works of Du Bois-Reymond 
\cite{DBR}, Veronese \cite{veronese, veronese2}, Levi-Civita \cite{LC} and
others \ref{hilb},\cite{Enriques1},\cite{Fano1}, nevertheless they were
fought (and defeated) by the likes of Russell (see e.g. \cite{Russell1}) and
Peano \cite{peano}. Their defeat was so complete that many of the
mathematicians of my generation even ignore the existence of the studies of
the above mentioned scholars. Probably, even in the world of science,
history is written only by the winners \cite{bair}. To understand, at least
in part, the cultural dynamics of this historical period in relation to the
infinitesimal, we suggest some works by Ehrlich \cite{el06}. Not much better
was the reception of the Non Standard Analysis created in the '60s by
Robinson \cite{rob66}, \cite{fle}, which arouses the contempt of (almost)
all those who do not know it, even though a minority of mathematicians of
the highest level has elaborated on it interesting theories (see e.g. \cite%
{A}, \cite{nelson}, \cite{tao}). A brief historical survey on NAM\ can be
found e.g. in \cite{BFreg}, \cite{Bla}.

Personally, I am convinced that the NAM is a very rich branch of mathematics
and allows to construct models of the real world in a more efficient way. In
this paper, we will show how NAM allows to construct a formalism for QM
which is closer to the ideas of Born, Heisenberg and Bohr than the formalism
of Von Neumann. This formalism is based on the theory of ultrafunctions
which are functions defined over a non Archimedean field, namely a field
which does not satisfy the Principle of Archimedes, and hence it contains
infinitesimal and infinite numbers.

\subsection{Ultrafunctions and Euclidean numbers}

The ultrafunctions can be considered as a kind generalized functions. In
many circumstances, the notion of real function is not sufficient to the
needs of a theory and it is necessary to extend it. The intensive use of the
Laplace transform in engineering led to the heuristic use of symbolic
methods, called operational calculus. An influential book on operational
calculus was Oliver Heaviside's Electromagnetic Theory of 1899 \cite{Hea}.
However his methods had a bad reputation among pure mathematicians because
they were not rigorous. When the Lebesgue integral was introduced, for the
first time a notion of generalized function became central to mathematics
since the notion of function was replaced by something defined almost
everywhere and not pointwise. During the late 1920s and 1930s further steps
were taken, very important to future work. The Dirac delta function was
boldly defined by Paul Dirac as a central aspect of his scientific
formalism. Jean Leray and Sergei Sobolev, working in partial differential
equations, defined the first adequate theory of generalized functions in
order to work with weak solutions of partial differential equations.
Sobolev's work was further developed in an extended form by Laurent Schwartz 
\cite{Schwartz}. To day, among people working in partial differential
equations, the theory of distributions of L. Schwartz is the most commonly
used, but also other notions of generalized functions have been introduced
by J.F. Colombeau \cite{col85} and M. Sato \cite{sa59}.

The ultrafunctions can be considered as a new kind of generalized functions.
They have been recently introduced \cite{ultra} and developed in \cite%
{belu2012,belu2013,milano,algebra,beyond,benci}. They provide generalized
solutions to certain equations which do not have any solution, not even
among the distributions. Actually, the ultrafunctions are pointwise defined
functions on a suitable subset of $\mathbb{E}^{N},$ and take their value in $%
\mathbb{E}$ where $\mathbb{E}$ is a non Archimedean field which contains the
real numbers. This fact allows to define the \textit{Dirac delta
ultrafunction }$\delta _{a}$\textit{\ }as a function which takes an infinite
value in the point $a$ and vanishes in the other points. So, in this
context, expression such as $\sqrt{\delta _{a}}$ or $\delta _{a}^{2}$ make
absolutely sense. The field $\mathbb{E},$ introduced in \cite{BF}, is called
field of Euclidean numbers and it is a particular field of hyperreal
numbers. We recall that the fields of hyperreal numbers are the basic fields
on which nonstandard analysis (NSA in the sequel) is based. In fact, in the
theory of ultrafunctions, a large use of the techniques of NSA is employed
even if sometimes it is hidden by the formalism which we have used. This
formalism is based on the notion of $\Lambda $-limit. Before ending this
introduction, we want to emphasize the differences by our approach to
nonstandard methods and the usual one: there are two main differences, one
in the aims and one in the methods.

Let us examine the difference in the aims. We think that infinitesimal and
infinite numbers should not be considered just as entities living in a
parallel universe (the nonstandard universe) which are only a tool to prove
some statement relative to our universe (the standard universe), but rather
that they should be considered mathematical entities which have the same
status of the others and can be used to build models as any other
mathematical entity. Actually, the advantages of a theory which includes
infinitesimals rely more on the possibility of making new models rather than
in the proving techniques. This paper, as well as \cite%
{BGG,ultra,milano,BHW,BBG,BLS}, are inspired by this principle.

As far as the methods are concerned, we introduce a non-Archimedean field
via a new notion of limit (see section \ref{OL}) and we us a language closer
to analysis and to applied Mathematics rather than to Logic.

\subsection{Notations\label{not}}

Let $\Omega $\ be an open subset of $\mathbb{R}^{N}$: then

\begin{itemize}
\item $C\left( \Omega \right) $ denotes the set of continuous functions
defined on $\Omega \subset \mathbb{R}^{N};$

\item $C_{c}\left( \Omega \right) $ denotes the set of continuous functions
in $C\left( \Omega \right) $ having compact support in $\Omega ;$

\item $C^{k}\left( \Omega \right) $ denotes the set of functions defined on $%
\Omega \subset \mathbb{R}^{N}$ which have continuous derivatives up to the
order $k;$

\item $C_{c}^{k}\left( \Omega \right) $ denotes the set of functions in $%
C^{k}\left( \Omega \right) \ $having compact support;

\item $L^{2}\left( \Omega \right) $ denotes the space of square integrable
functions defined almost everywhere in $\Omega $;

\item $\mathfrak{mon}(x)=\{y\in \mathbb{E}^{N}\ |\ x\sim y\}\ $(see Def. \ref%
{MG});

\item $\mathfrak{gal}(x)=\{y\in \mathbb{E}^{N}\ |\ x-y$ is a finite number$%
\}\ $(see Def. \ref{MG});

\item given any set $E\subset X$, $\chi _{E}:X\rightarrow \mathbb{R}$
denotes the characteristic function of $E$, namely%
\begin{equation*}
\chi _{E}(x):=\left\{ 
\begin{array}{cc}
1 & if\ \ x\in E \\ 
&  \\ 
0 & if\ \ x\notin E%
\end{array}%
\right.
\end{equation*}

\item with some abuse of notation we set $\chi _{a}(x):=\chi _{\left\{
a\right\} }(x);$

\item $\partial _{i}=\frac{\partial }{\partial x_{i}}$ denotes the usual
partial derivative; $D_{i}$ denotes the generalized derivative (see section %
\ref{adu});

\item $\int $ denotes the usual Lebesgue integral; $\doint $ denotes the
pointwise integral (see section \ref{adu}).

\item if $E$ is any set, then $|E|$ denotes its cardinality.
\end{itemize}

\section{$\Lambda $-theory\label{lt}}

As we have already remarked in the introduction, $\Lambda $-theory can be
considered a different approach to Nonstandard Analysis\footnote{%
In \cite{fle} and \cite{BDNF2}, the reader can find several other approaches
to NSA and an analysis of them.}. It can be introduced via the notion of $%
\Lambda $-limit, and it can be easily used for the purposes of this paper.
We introduce the Euclidean numbers (which are the basic object of $\Lambda $%
-theory) via an algebraic approach as in \cite{benci99}. An elementary
presentation of (part of) this theory can be found in \cite{Bmathesis} and 
\cite{bencilibro}.

\subsection{Non Archimedean Fields\label{naf}}

Here, we recall the basic definitions and facts regarding Non Archimedean
fields. In the following, ${\mathbb{K}}$ will denote an ordered field. We
recall that such a field contains (a copy of) the rational numbers. Its
elements will be called numbers.

\begin{definition}
Let $\mathbb{K}$ be an ordered field. Let $\xi \in \mathbb{K}$. We say that:

\begin{itemize}
\item $\xi $ is infinitesimal if, for all positive $n\in \mathbb{N}$, $|\xi
|<\frac{1}{n}$;

\item $\xi $ is finite if there exists $n\in \mathbb{N}$ such that $|\xi |<n$%
;

\item $\xi $ is infinite if, for all $n\in \mathbb{N}$, $|\xi |>n$
(equivalently, if $\xi $ is not finite).
\end{itemize}
\end{definition}

\begin{definition}
An ordered field $\mathbb{K}$ is called Non-Archimedean if it contains an
infinitesimal $\xi \neq 0$.
\end{definition}

It's easily seen that all infinitesimal are finite, that the inverse of an
infinite number is a nonzero infinitesimal number, and that the inverse of a
nonzero infinitesimal number is infinite.

\begin{definition}
A superreal field is an ordered field $\mathbb{K}$ that properly extends $%
\mathbb{R}$.
\end{definition}

It is easy to show that any superreal field contains infinitesimal and
infinite numbers. Infinitesimal numbers can be used to formalize a new
notion of "closeness":

\begin{definition}
\label{def infinite closeness} We say that two numbers $\xi ,\zeta \in {%
\mathbb{K}}$ are infinitely close if $\xi -\zeta $ is infinitesimal. In this
case we write $\xi \sim \zeta $.
\end{definition}

Clearly, the relation "$\sim $" of infinite closeness is an equivalence
relation.

\begin{theorem}
If $\mathbb{K}$ is a superreal field, every finite number $\xi \in \mathbb{K}
$ is infinitely close to a unique real number $r\sim \xi $, called the the 
\textbf{standard part} of $\xi $.
\end{theorem}

Given a finite number $\xi $, we denote it standard part by $st(\xi )$, and
we put $st(\xi )=+\infty $ ($st(\xi )=-\infty $) if $\xi \in \mathbb{K}$ is
a positive (negative) infinite number.

\begin{definition}
\label{MG}Let $\mathbb{K}$ be a superreal field, and $\xi \in \mathbb{K}$ a
number. The \label{def monad} monad of $\xi $ is the set of all numbers that
are infinitely close to it:%
\begin{equation*}
\mathfrak{m}\mathfrak{o}\mathfrak{n}(\xi )=\{\zeta \in \mathbb{K}:\xi \sim
\zeta \},
\end{equation*}%
and the galaxy of $\xi $ is the set of all numbers that are finitely close
to it: 
\begin{equation*}
\mathfrak{gal}(\xi )=\{\zeta \in \mathbb{K}:\xi -\zeta \ \text{is\ finite}\}.
\end{equation*}
\end{definition}

By definition, it follows that the set of infinitesimal numbers is $%
\mathfrak{mon}(0)$ and that the set of finite numbers is $\mathfrak{gal}(0)$%
. Moreover, the standard part can be regarded as a function:%
\begin{equation}
st:\mathfrak{gal}(0)\rightarrow \mathbb{R}.  \label{sh}
\end{equation}

Any ordered field $\mathbb{K}$ can be complexified to get a new field 
\begin{equation*}
\mathbb{K}+i\mathbb{K}
\end{equation*}%
namely a field of numbers of the form 
\begin{equation*}
a+ib,\ a,b\in \mathbb{K}.
\end{equation*}%
Thus also the complexification of non-Archimedean fields does not present
any particular peculiarity.

\subsection{The Euclidean numbers\label{OL}}

Let $\Lambda $ be an infinite set containing $\mathbb{R}$ and let $\mathfrak{%
L}$ be the family of finite subsets of $\Lambda .$ A function $\varphi :%
\mathfrak{L}\rightarrow E$ will be called \textit{net }(with values in $E$).
The set of such nets is denoted by $\mathfrak{F}\left( \mathfrak{L},\mathbb{R%
}\right) .$ Such a set is a real algebra equipped with the natural operations%
\begin{eqnarray*}
\left( \varphi +\psi \right) (\lambda ) &=&\varphi (\lambda )+\psi (\lambda
); \\
\left( \varphi \cdot \psi \right) (\lambda ) &=&\varphi (\lambda )\cdot \psi
(\lambda );
\end{eqnarray*}%
and the partial order relation:%
\begin{equation*}
\varphi \geq \psi \Leftrightarrow \forall \lambda \in \mathfrak{L,\ }\varphi
(\lambda )\geq \psi (\lambda ).
\end{equation*}

\begin{definition}
The set of Euclidean numbers $\mathbb{E\supset R}$ is a field such that
there is a surjective homomorphism 
\begin{equation*}
J:\mathfrak{F}\left( \mathfrak{L},\mathbb{R}\right) \rightarrow \mathbb{E}
\end{equation*}%
or, more exactly a map which satisfies the following properties:

\begin{itemize}
\item $J\left( \varphi +\psi \right) =J\left( \varphi \right) +J\left( \psi
\right) ;$

\item $J\left( \varphi \cdot \psi \right) =J\left( \varphi \right) \cdot
J\left( \psi \right) ;$

\item \textit{if }$\varphi (\lambda )\geq r,$ then $J\left( \varphi \right)
\geq r.$
\end{itemize}
\end{definition}

The proof of the existence of such a field is an easy consequence of the
Krull-Zorn theorem. It can be found, e.g. in \cite%
{benci99,ultra,milano,bencilibro}. In this paper, we use also the
complexification of $\mathbb{E}$, denoted by%
\begin{equation}
\mathbb{C}^{\ast }=\mathbb{E}+i\mathbb{E}.  \label{rita}
\end{equation}

The number $J\left( \varphi \right) $ is called the $\Lambda $ limit of the
net $\varphi $ and will be denoted by%
\begin{equation*}
J\left( \varphi \right) =\lim_{\lambda \uparrow \Lambda }\varphi (\lambda )
\end{equation*}%
The reason of this name/notation is that the operation%
\begin{equation*}
\varphi \mapsto \lim_{\lambda \uparrow \Lambda }\varphi (\lambda )
\end{equation*}%
satisfies many of the properties of the usual limit, more exactly it
satisfies the following properties:

\begin{itemize}
\item \textsf{(}$\Lambda $-\textsf{1)}\ \textbf{Existence.}\ \textit{Every
net }$\varphi :\mathfrak{L}\rightarrow \mathbb{R}$\textit{\ has a unique
limit }$L\in \mathbb{E}{.}$

\item ($\Lambda $-2)\ \textbf{Constant}. \textit{If }$\varphi (\lambda )$%
\textit{\ is} \textit{eventually} \textit{constant}, \textit{namely} $%
\exists \lambda _{0}\in \mathfrak{L},r\in \mathbb{R}$ such that $\forall
\lambda \supset \lambda _{0},\ \varphi (\lambda )=r,$ \textit{then}%
\begin{equation*}
\lim_{\lambda \uparrow \Lambda }\varphi (\lambda )=r.
\end{equation*}

\item ($\Lambda $-3)\ \textbf{Sum and product}.\ \textit{For all }$\varphi
,\psi :\mathfrak{L}\rightarrow \mathbb{R}$\emph{: }%
\begin{eqnarray*}
\lim_{\lambda \uparrow \Lambda }\varphi (\lambda )+\lim_{\lambda \uparrow
\Lambda }\psi (\lambda ) &=&\lim_{\lambda \uparrow \Lambda }\left( \varphi
(\lambda )+\psi (\lambda )\right) ; \\
\lim_{\lambda \uparrow \Lambda }\varphi (\lambda )\cdot \lim_{\lambda
\uparrow \Lambda }\psi (\lambda ) &=&\lim_{\lambda \uparrow \Lambda }\left(
\varphi (\lambda )\cdot \psi (\lambda )\right) .
\end{eqnarray*}
\end{itemize}

Now let us see the main differences between the usual limit (which we will
call Cauchy limit) and the $\Lambda $-limit. We recall the definition of
Cauchy limit (as formalized by Weierstrass):%
\begin{equation*}
L=\lim_{\lambda \rightarrow \Lambda }\varphi (\lambda )
\end{equation*}%
if and only if, $\forall \varepsilon \in \mathbb{R}^{+}$, $\exists \lambda
_{0}\in \mathfrak{L},$ such that $\forall \lambda \supset \lambda _{0},\ $%
\begin{equation*}
|\varphi (\lambda )-L|\ \leq \varepsilon
\end{equation*}%
The classical example of Cauchy limit of a net is provided by the definition
of the Cauchy integral:%
\begin{equation*}
\int_{a}^{b}f(x)dx=\lim_{\lambda \rightarrow \Lambda }\sum_{x\in \left[ a,b%
\right] \cap \lambda }f(x)(x^{+}-x);\ \ x^{+}=\min \left\{ y\in \mathbb{R}%
\cap \lambda \ |\ y>x\right\}
\end{equation*}

Notice that in order to distinguish the two kind of limits we have used the
symbols "$\lambda \uparrow \Lambda $" and "$\lambda \rightarrow \Lambda $"
respectively. Since also the Cauchy limit (when it exists) satisfies ($%
\Lambda $-2) and ($\Lambda $-3) the only difference between the the Cauchy
limit and the $\Lambda $-limit is that the latter always exists. This fact
implies that $\mathbb{E}$ must be larger than $\mathbb{R}$ since otherwise a
diverging net cannot have a limit. In the case in which the Cauchy limit
exists the relation between the two limits is given by the following
identity:%
\begin{equation}
\lim_{\lambda \rightarrow \Lambda }\varphi (\lambda )=st\left( \lim_{\lambda
\uparrow \Lambda }\varphi (\lambda )\right)  \label{bn}
\end{equation}

In order to give a feeling of the Euclidean number, we will describe a
possible interpretation of some of them. If $E\subset \mathbb{R}\subset
\Lambda ,$ we set%
\begin{equation*}
\mathfrak{n}\left( E\right) =\lim_{\lambda \uparrow \Lambda }\ |E\cap
\lambda |
\end{equation*}%
where $|F|$ denotes the number of elements of the finite set $F$. Notice
that $E\cap \lambda $ is a finite set since $\lambda \in \mathfrak{L.}$ Then
the above limit makes sense since for every $\lambda \in \mathfrak{L}$, $%
|E\cap \lambda |~\in \mathbb{N}\subset \mathbb{R}$. If $E$ is a finite set,
the sequence is eventually constant, namely, $\forall \lambda \supset E,\
|E\cap \lambda |~=|E|$ and hence $\mathfrak{n}\left( E\right) =|E|$. If $E$
is an infinite set, the above limit gives an infinite number. Hence $%
\mathfrak{n}\left( E\right) $ extends the "measure of the size of a set" to
infinite sets. The Euclidean number $\mathfrak{n}\left( E\right) $ is called 
\textit{numerosity} of $E$. For example, the number $\alpha $ defined by%
\begin{equation}
\alpha :=\lim_{\lambda \uparrow \Lambda }\ |\mathbb{N}\cap \lambda |
\label{alfa}
\end{equation}%
is a measure of the size of $\mathbb{N}=\left\{ 1,2,3,...\right\} $. The
theory of numerosity can be considered as an extension of the Cantorian
theory of cardinal and ordinal numbers. The reader interested to the details
and the developments of this theory is referred to \cite%
{benci95b,BDN2003,BDNF1,BF}.

\subsection{Extension of functions and grid functions\label{HE}}

If $\varphi (\lambda )=\left( \varphi _{1}(\lambda ),...,\varphi
_{N}(\lambda )\right) \in \mathbb{R}^{N},$ we set%
\begin{equation*}
\lim_{\lambda \uparrow \Lambda }\varphi (\lambda )=\left( \lim_{\lambda
\uparrow \Lambda }~\varphi _{1}(\lambda ),...,\lim_{\lambda \uparrow \Lambda
}~\varphi _{N}(\lambda )\right) \in \mathbb{E}^{N}
\end{equation*}%
Given a set $A\subset \mathbb{R}^{N}$, we define 
\begin{equation*}
A^{\ast }=\left\{ \lim_{\lambda \uparrow \Lambda }\varphi (\lambda )\ |\
\forall \lambda ,\ \varphi (\lambda )\in A\right\} ;
\end{equation*}%
following Keisler \cite{keisler76}, $A^{\ast }$ will be called the \textbf{%
natural extension} of $A.$ Clearly we have that $\mathbb{R}^{\ast }=\mathbb{E%
}$. This fact justifies the notation (\ref{rita}) to denote the
complexification of $\mathbb{E}$: 
\begin{equation*}
\mathbb{C}^{\ast }=\mathbb{E}+i\mathbb{E}=\mathbb{R}^{\ast }+i\mathbb{R}%
^{\ast }.
\end{equation*}%
Any function 
\begin{equation*}
f:A\rightarrow \mathbb{R},\ \ A\subset \mathbb{R}^{N}
\end{equation*}%
can be univocally extended to $A^{\ast }$ by setting 
\begin{equation*}
f^{\ast }\left( \lim_{\lambda \uparrow \Lambda }\ x_{\lambda }\right)
=\lim_{\lambda \uparrow \Lambda }~f^{\ast }\left( x_{\lambda }\right) ;
\end{equation*}%
the function 
\begin{equation*}
f^{\ast }:A^{\ast }\rightarrow \mathbb{E},
\end{equation*}%
will be called \textbf{natural extension} of $f.$ More in general, if%
\begin{equation*}
u_{\lambda }:A\rightarrow \mathbb{R},\ \ A\subset \mathbb{R}^{N}
\end{equation*}%
is a net of functions, we define the $\Lambda $-limit 
\begin{equation*}
u=\lim_{\lambda \uparrow \Lambda }~u_{\lambda }:A^{\ast }\rightarrow \mathbb{%
E},
\end{equation*}%
as follows: for any $x=\lim_{\lambda \uparrow \Lambda }\ x_{\lambda }\in 
\mathbb{E}^{N}$, we set 
\begin{equation*}
u(x)=\lim_{\lambda \uparrow \Lambda }~u_{\lambda }\left( x_{\lambda }\right) 
\end{equation*}%
In particular, if, for all $x\in \mathbb{R}^{N}$ 
\begin{equation*}
v(x)=\lim_{\lambda \rightarrow \Lambda }~u_{\lambda }\left( x\right) 
\end{equation*}%
by (\ref{bn}), it follows that%
\begin{equation*}
\forall x\in \mathbb{R}^{N},\ \ v(x)=st\left[ u(x)\right] .
\end{equation*}

\begin{definition}
We say that a set $F\subset \mathbb{E}$ is \textbf{hyperfinite} if there is
a net $\left\{ F_{\lambda }\right\} _{\lambda \in \Lambda }$ of finite set
such that 
\begin{equation*}
F=\left\{ \lim_{\lambda \uparrow \Lambda }\ x_{\lambda }\ |\ x_{\lambda }\in
F_{\lambda }\right\}
\end{equation*}
\end{definition}

The hyperfinite sets share many properties of finite sets. For example, it
is possible to "add" the elements of an hyperfinite set of numbers. If $F$
is an hyperfinite set of numbers, the \textbf{hyperfinite sum} of the
elements of $F$ is defined in the following way: 
\begin{equation*}
\sum_{x\in F}x=\ \lim_{\lambda \uparrow \Lambda }\sum_{x\in F_{\lambda }}x.
\end{equation*}

\begin{definition}
A hyperfinite set $\Gamma $ such that $\mathbb{R}^{N}\subset \Gamma \subset 
\mathbb{E}^{N}$ is called \textbf{hyperfinite grid}.
\end{definition}

If $\Gamma _{\lambda }$ is a family of finite subsets of of $\mathbb{R}$
which satisfy the following property:%
\begin{equation*}
\mathbb{R}^{N}\cap \lambda \subset \Gamma _{\lambda }
\end{equation*}%
it is not difficult to prove that the the set%
\begin{equation*}
\Gamma =\left\{ \lim_{\lambda \uparrow \Lambda }\ x_{\lambda }\ |\
x_{\lambda }\in \Gamma _{\lambda }\right\} 
\end{equation*}%
is a hyperfinite grid. From now on $\Gamma $ will denote a hyperfinite grid
fixed once forever.

\begin{definition}
A space of grid functions is a family $\mathfrak{G}(\Gamma )$ of functions 
\begin{equation*}
u:\Gamma \rightarrow \mathbb{R}
\end{equation*}%
such that, for every $x=\ \lim_{\lambda \uparrow \Lambda }x_{\lambda }\in
\Gamma $, we have that 
\begin{equation*}
u(x)=\ \lim_{\lambda \uparrow \Lambda }\ u_{\lambda }(x_{\lambda }).
\end{equation*}
\end{definition}

If $f\in \mathfrak{F}(\mathbb{R}^{N})$, and $x=\ \lim_{\lambda \uparrow
\Lambda }x_{\lambda }\in \Gamma $, we set 
\begin{equation}
f%
{{}^\circ}%
(x):=\lim_{\lambda \uparrow \Lambda }\ f(x_{\lambda })  \label{lina}
\end{equation}%
namely $f%
{{}^\circ}%
$ is the restriction to $\Gamma $ of the natural extension $f^{\ast }$ which
is defined on all $\mathbb{E}^{N}$.

It is easy to check that, for every $a\in \Gamma ,$ $\chi _{a}(x)$ is a grid
function, and hence every grid function can be represented by the following
hyperfinite sum:%
\begin{equation}
u(x)=\sum_{a\in \Gamma }u(a)\chi _{a}(x)  \label{pu}
\end{equation}%
namely $\left\{ \chi _{a}(x)\right\} _{a\in \Gamma }$ is a basis for $%
\mathfrak{G}(\mathbb{R}^{N})$. If a function, such as $1/|x|$ is not defined
in some point, we put $(1/|x|)%
{{}^\circ}%
$ equal to $0$ for $x=0;$ in general, if $E$ is a subset of $\mathbb{R}^{N}$
and $f$ is defined on $E$, we set%
\begin{equation*}
f%
{{}^\circ}%
(x)=\sum_{a\in \Gamma \cap E^{\ast }}f^{\ast }(a)\chi _{a}(x).
\end{equation*}%
Before ending this section we need an other definition. Given two function
spaces $V$ and $W,$ we set%
\begin{eqnarray*}
V^{\ast } &=&\left\{ \lim_{\lambda \uparrow \Lambda }\ u_{\lambda }\ |\
u_{\lambda }\in V\right\} \\
W^{\ast } &=&\left\{ \lim_{\lambda \uparrow \Lambda }\ u_{\lambda }\ |\
u_{\lambda }\in W\right\}
\end{eqnarray*}

\begin{definition}
\label{oper}An operator%
\begin{equation*}
F:V^{\ast }\longrightarrow W^{\ast }
\end{equation*}%
is called \textbf{internal} if there exists a net of operators $\left\{
F_{\lambda }\right\} _{\lambda \in \mathfrak{L}},$ such that%
\begin{equation*}
Fu:=\lim_{\lambda \uparrow \Lambda }\ F_{\lambda }\left( u_{\lambda }\right)
\end{equation*}%
where $u=\ \lim_{\lambda \uparrow \Lambda }u_{\lambda },$ $u_{\lambda }\in
V. $
\end{definition}

In general any mathematical entity is called "internal" if it is the $%
\Lambda $-limit of some other entities. In this section we have introduced
internal function (e.g. the grid functions, internal sets, hyperfinite sets
and internal operators). We do not need to develop the full theory which is
a basic tool in NSA. We have just introduced explicitly the objects which
are needed in this exposition.

\section{Ultrafunctions\label{Ultrafunctions}}

\subsection{Axiomatic definition of ultrafunctions\label{adu}}

Let $V=C_{c}(\mathbb{R}^{N})$ denote the space of continuous functions with
compact support. We will denote by $\left\{ V_{\lambda }\right\} _{\lambda
\in \mathfrak{L}}$ a directed set of all finite dimensional subspaces of $%
V(\Omega ),$ namely for every couple of spaces $V_{\lambda _{1}},V_{\lambda
_{2}}$ there exists $\lambda _{3}\supseteq \lambda _{1}\cup \lambda _{2}$
such that%
\begin{equation*}
V_{\lambda _{1}}+V_{\lambda _{2}}\subset V_{\lambda _{3}}.
\end{equation*}

A space of ultrafunctions $V%
{{}^\circ}%
$ modelled on $\left\{ V_{\lambda }\right\} _{\lambda \in \mathfrak{L}}$ is
a family of grid functions 
\begin{equation*}
u:\Gamma \rightarrow \mathbb{E}
\end{equation*}%
equipped with an internal functional%
\begin{equation*}
\doint :V%
{{}^\circ}%
\rightarrow \mathbb{E}
\end{equation*}%
(called \textbf{pointwise integral}) and an $N$ internal operators 
\begin{equation*}
D_{i}:V%
{{}^\circ}%
\rightarrow V%
{{}^\circ}%
\end{equation*}%
(called \textbf{generalized partial derivative}) which satisfy the following
axioms:

\begin{axiom}
\label{1}For any $u\in V%
{{}^\circ}%
,$ there exists a net $u_{\lambda }$ such that%
\begin{equation*}
u_{\lambda }\in V_{\lambda }
\end{equation*}%
and 
\begin{equation*}
u=\ \lim_{\lambda \uparrow \Lambda }\ u_{\lambda }.
\end{equation*}
\end{axiom}

\begin{axiom}
\label{2}If $u=\ \lim_{\lambda \uparrow \Lambda }u_{\lambda },$ $u_{\lambda
}\in V_{\lambda },$ then%
\begin{equation}
\doint u(x)dx=\lim_{\lambda \uparrow \Lambda }\ \dint u_{\lambda }(x)dx.
\label{int}
\end{equation}
\end{axiom}

\begin{axiom}
\label{3}If $a\in \Gamma $,%
\begin{equation*}
\doint \chi _{a}(x)dx>0.
\end{equation*}
\end{axiom}

\begin{axiom}
\label{4}If $u=\ \lim_{\lambda \uparrow \Lambda },$ $u_{\lambda }\in
V_{\lambda }\cap C^{1}(\mathbb{R}^{N}),$ and and $x=\ \lim_{\lambda \uparrow
\Lambda }x_{\lambda }\in \Gamma $, then%
\begin{equation}
D_{i}u(x)=\lim_{\lambda \uparrow \Lambda }\ \partial _{i}u_{\lambda
}(x_{\lambda }).  \label{der}
\end{equation}
\end{axiom}

\begin{axiom}
\label{5}For every $u\in V%
{{}^\circ}%
$%
\begin{equation*}
D_{i}u=0\Leftrightarrow u=0.
\end{equation*}
\end{axiom}

\begin{axiom}
If we set 
\begin{equation*}
\mathfrak{supp}\left( u\right) =\left\{ x\in \Gamma \ |\ u(x)\neq 0\right\} ,
\end{equation*}%
then%
\begin{equation*}
\mathfrak{supp}\left( D_{i}\chi _{a}(x)\right) \subset \mathfrak{mon}(a).
\end{equation*}
\end{axiom}

\begin{axiom}
\label{7}For every $u,v\in V%
{{}^\circ}%
$%
\begin{equation}
\doint D_{i}u(x)v(x)dx=-\doint u(x)D_{i}v(x)dx.  \label{parti}
\end{equation}
\end{axiom}

In the literature several spaces of ultrafunctions have been introduced and
developed (see  \cite{ultra, belu2012, belu2013, milano, algebra, beyond}
and the references therein). However the proof of a model of ultrafunctions
which satisfies \textbf{all} the above seven axioms is a delicate matter and
we refer to \cite{benci}.

Now, we will briefly discuss these axioms. The first axiom characterizes the
ultrafunctions with respect to other internal functions. The second axiom is
nothing else but the definition of the pointwise integral. By its
definition, for every function $f\in V$, 
\begin{equation*}
\doint f%
{{}^\circ}%
(x)dx=\int f(x)dx
\end{equation*}%
Then it extends the usual Riemann integral from $V=C_{0}(\mathbb{R}^{N})$ to 
$V%
{{}^\circ}%
.$ Axiom $3$ shows that the above inequality cannot hold for all the Riemann
integrable function since%
\begin{equation*}
\int \chi _{a}(x)dx=0\neq \doint \chi _{a}(x)dx
\end{equation*}%
This Axiom is natural, since when we work in a non-Archimedan world the
infinitesimals matter and cannot be forgotten as the Riemann integral does.
Also the above inequality shows that it is necessary to use a different
symbol (namely $\doint $) to distinguish the pointwise integral from the
Riemann or the Lebesgue integral. Axiom $4$ shows that the generalized
derivative extends the usual derivative; in fact if $f$ $\in C^{1}(\mathbb{R}%
^{N})$ and $x\in \mathbb{R}^{N},$ then%
\begin{equation*}
D_{i}f%
{{}^\circ}%
(x)=\lim_{\lambda \uparrow \Lambda }\ \partial _{i}f(x)=\partial _{i}f(x).
\end{equation*}%
On the other hand the operator $D_{i}$ is defined on all the functions and
it must be defined in such a way that the most useful property of the usual
derivative be satisfied, and this is the content of the last three axiom.
Axiom $5$ says that the ultrafunctions behave as compactly supported $C^{1}$
functions. Axiom $6$ states that the derivative is a local operator. Axiom $7
$ states a formula which is of primary importance in the theory of weak
derivatives, distribution, calculus of variations etc. Usually this formula
is deduced by the Leibniz rule%
\begin{equation*}
D(fg)=Dfg+fDg
\end{equation*}%
However, the Leibniz rule cannot be satisfied by every ultrafunction by the
Schwartz impossibility theorem (see \cite{Schwartz},\cite{algebra}; see also
the footnote at p. \pageref{leib}). Nevertheless the identity (\ref{parti})
holds for all the ultrafunctions.

\subsection{Structure of the space of ultrafunctions}

By (\ref{pu}) and Axiom \ref{3}, it follows that, for every $u\in V%
{{}^\circ}%
,$%
\begin{equation*}
\doint u(x)dx=\dsum\limits_{a\in \Gamma }u(a)d\left( a\right) 
\end{equation*}%
where%
\begin{equation*}
\forall a\in \Gamma ,\ d(a)=\doint \chi _{a}(x)dx,
\end{equation*}%
namely the pointwise integral reduces to a hyperfinite sum. This fact
justifies its name.

In view of our application to QM, from now on, we will consider complex
valued ultrafunctions, namely ultrafunctions in the space%
\begin{equation*}
H%
{{}^\circ}%
:=V%
{{}^\circ}%
\oplus iV%
{{}^\circ}%
\end{equation*}%
Also we shall use the notations%
\begin{equation*}
H:=V\oplus iV=C_{c}\left( \mathbb{R}^{N},\mathbb{C}\right)
\end{equation*}%
\begin{equation*}
H_{\lambda }:=V_{\lambda }\oplus iV_{\lambda }
\end{equation*}

The pointwise integral allows to define the following sesquilinear form on $H%
{{}^\circ}%
$: 
\begin{equation}
\doint u(x)\overline{v(x)}dx=\sum_{x\in \Gamma }u(x)\overline{v(x)}d(x).
\label{rina}
\end{equation}%
Here $\overline{z}$ represents the complex conjugate of $z$. By virtue of
Axiom \ref{3}, such a form is scalar product. The norm of an ultrafunction
will be given by 
\begin{equation*}
\left\Vert u\right\Vert =\left( \doint |u(x)|^{2}\ dx\right) ^{\frac{1}{2}}.
\end{equation*}%
Also, the pointwise integral allows us to define the \textbf{delta (or the
Dirac) ultrafunction} as follows: for every $a\in \Gamma $,%
\begin{equation*}
\delta _{a}(x)=\frac{\chi _{a}(x)}{d(a)}.
\end{equation*}%
In fact, for every $u\in V%
{{}^\circ}%
,$ we have that%
\begin{equation*}
\doint \delta _{a}(x)u(x)dx=\dsum\limits_{x\in \Gamma }u(x)\delta
_{a}(x)d(x)=\dsum\limits_{x\in \Gamma }u(x)\frac{\chi _{a}(x)}{d(a)}d(x)=u(a)
\end{equation*}

The delta functions are orthogonal with each other with respect to the
scalar product (\ref{rina}); hence, if normalized, they provide an
orthonormal basis, called \textbf{delta-basis}, given by%
\begin{equation}
\left\{ \delta _{a}\sqrt{d(a)}\right\} _{a\in \Gamma }=\left\{ \frac{\chi
_{a}}{\sqrt{d(a)}}\right\} _{a\in \Gamma }  \label{db}
\end{equation}%
Hence, every ultrafunction can be represented as follows:%
\begin{equation}
u(x)=\sum_{a\in \Gamma }\left( \doint u(\xi )\delta _{a}(\xi )d\xi \right)
\chi _{a}(x).  \label{eq:lella}
\end{equation}

The scalar product allows the following proposition:

\begin{proposition}
\label{dual}If 
\begin{equation*}
\Phi :H\rightarrow \mathbb{C}
\end{equation*}%
is a linear internal functional, then there exists $u_{\Phi }$ such that,
for all $v=\lim_{\lambda \uparrow \Lambda }v_{\lambda }\in H%
{{}^\circ}%
$%
\begin{equation*}
\doint u_{\Phi }v~dx=\lim_{\lambda \uparrow \Lambda }\ \Phi \left(
v_{\lambda }\right)
\end{equation*}%
and for every $f\in V,$%
\begin{equation*}
\doint u_{\Phi }f%
{{}^\circ}%
dx=\ \Phi \left( f\right)
\end{equation*}
\end{proposition}

\textbf{Proof}: If $v\in H_{\lambda },$ then the map%
\begin{equation*}
v\mapsto \Phi \left( v\right)
\end{equation*}%
is a linear functional over $H_{\lambda }$ and hence, since there exists $%
u_{\lambda }\in H_{\lambda }$ such that $\forall v\in H_{\lambda },$%
\begin{equation*}
\int u_{\lambda }v~dx=\Phi \left( v\right) .
\end{equation*}%
If we set%
\begin{equation*}
u_{\Phi }=\lim_{\lambda \uparrow \Lambda }u_{\lambda }
\end{equation*}%
the conclusion follows.

$\square $

If we want to apply the theory of ultrafuctions to QM, one of the most
important thing to analyze is their relation to the $L^{2}$-functions; in
fact in the usual formalism of QM a physical state is described by a $L^{2}$%
-function $\psi $, but we cannot associate an ultrafunction $u$ to a
function $\psi $ by using eq.(\ref{lina}) since the $L^{2}$-functions are
not pointwise defined. Then we need a new definition.

\begin{definition}
\label{L2}Given a function $\psi \in L^{2}\left( \Omega \right) ,$ we denote
by $\psi 
{{}^\circ}%
$ the unique ultrafunction such that, for every $v=\lim_{\lambda \uparrow
\Lambda }v_{\lambda }(x)\in H%
{{}^\circ}%
$, 
\begin{equation*}
\doint \psi 
{{}^\circ}%
v~dx=\lim_{\lambda \uparrow \Lambda }\int_{\Omega }\psi v_{\lambda }dx.
\end{equation*}
\end{definition}

The above definition is well posed since the map%
\begin{equation*}
\Phi :v\mapsto \int \psi vdx
\end{equation*}%
is a functional on the space $H%
{{}^\circ}%
$ an hence, by Prop. \ref{dual}, there exists an ultrafunction $\psi 
{{}^\circ}%
$ such that%
\begin{equation*}
\doint \psi 
{{}^\circ}%
v~dx=\Phi \left( v\right) .
\end{equation*}

\subsection{Self-adjoint operators on ultrafunctions}

If 
\begin{equation*}
L:H%
{{}^\circ}%
\rightarrow H%
{{}^\circ}%
,\ \ \ H%
{{}^\circ}%
=V%
{{}^\circ}%
\oplus iV%
{{}^\circ}%
\end{equation*}%
is an internal linear operator, it can be regarded as a infinite matrix
since, by Def. \ref{oper},%
\begin{equation*}
Lu:=\lim_{\lambda \uparrow \Lambda }\ L_{\lambda }u_{\lambda }
\end{equation*}%
where%
\begin{equation*}
L_{\lambda }:H_{\lambda }\rightarrow H_{\lambda };\ \ H_{\lambda
}=V_{\lambda }\oplus iV_{\lambda }
\end{equation*}%
can be represented by a matrix since $H_{\lambda }$ is a finite dimensional
space over $\mathbb{C}$. Then if $L$ is a selfadjoint operator, namely 
\begin{equation*}
\doint Lu\overline{v}~dx=\doint u\overline{Lv}~dx
\end{equation*}%
the matrices $L_{\lambda }$ are Hermitian. $L$ can be regarded as an
infinite dimensional Hermitian matrix. Hence $\sigma (L),$ the spectrum of $L
$ consists of eigenvalues only, more exactly%
\begin{equation*}
\sigma (L)=\left\{ \lim_{\lambda \uparrow \Lambda }\ \mu _{\lambda }\in 
\mathbb{E}\ |\ \forall \lambda ,\mu _{\lambda }\in \sigma (L_{\lambda
})\right\} .
\end{equation*}%
The corresponding normalized eigenfunctions form an orthonormal basis of $V%
{{}^\circ}%
$. So, in the ultrafunction approach, as in the finite dimensional
vector-spaces, the distinction between self-adjoint operators and Hermitian
operators is not needed since every Hermitian operator is self-adjoint.

Now, let us analyze the main selfadjoint/Hermitian operators of QM in the
frame of ultrafunctions. The \textbf{position operator}%
\begin{equation*}
\mathbf{q}:V%
{{}^\circ}%
\rightarrow \left( V%
{{}^\circ}%
\right) ^{N}
\end{equation*}%
is defined by%
\begin{equation}
\left( \mathbf{q}u\right) \left( x\right) =xu(x).  \label{qu}
\end{equation}%
It is immediate to check that $\sigma (\mathbf{p})=\Gamma $ and that the
corresponding orthonormal basis is given by the $\delta $-basis (\ref{db});
in fact%
\begin{equation*}
\left( \mathbf{q}\delta _{q}\right) \left( x\right) =x\delta _{q}(x)=q\delta
_{q}(x)
\end{equation*}%
since for $x\neq q$, $\delta _{q}(x)=0.$

The \textbf{momentum operator}%
\begin{equation*}
\mathbf{p}:V%
{{}^\circ}%
\rightarrow \left( V%
{{}^\circ}%
\right) ^{N}
\end{equation*}%
is defined by%
\begin{equation}
\left( \mathbf{p}u\right) \left( x\right) =-iDu(x)=-i\left(
D_{1}u(x),...,D_{N}u(x)\right) \footnote{%
Here and in the rest of this paper, we assume $\hbar =1$.}  \label{pi}
\end{equation}%
Its spectrum $\sigma (\mathbf{p})$ cannot be computed explicitly, but it is
possible to prove that infinitely close to any vector $v\in \mathbb{R}^{N}$
there is at least an eigenvalue (actually infinitely many eigenvalues) $k\in
\sigma (\mathbf{p})$: $k\sim v$. The corresponding orthonormal basis is
given by%
\begin{equation}
\left\{ \frac{e^{ik\cdot x}}{\sqrt{\beta }}\right\} _{\mathbf{k}\in \sigma (%
\mathbf{p})}  \label{perla}
\end{equation}%
where $\sqrt{\beta }$ is a normalization factor, namely%
\begin{equation*}
\beta =\doint \left\vert e^{ik\cdot x}\right\vert ~dx=\doint dx=\sum_{x\in
\Gamma }d(x)
\end{equation*}%
is an infinite number. Notice that $e^{ik\cdot x}$ is an eigenvalue for the
operator $-iD$ only for some particular values of $k\in \mathbb{E}^{N}$. If
you take $k\in \mathbb{R}^{N}$ arbitrarily, it might happen that $\left(
e^{ik\cdot x}\right) 
{{}^\circ}%
$ is not an eigenfunction of $-iD,$ since the equality%
\begin{equation*}
D_{i}f%
{{}^\circ}%
=\left( \partial _{i}f\right) 
{{}^\circ}%
,
\end{equation*}%
holds only for functions $f\in V\cap C^{1}(\mathbb{R}^{N})=C_{c}^{1}(\mathbb{%
R}^{N})$ (see Axiom \ref{4}). So, for a generic $k$ the equality%
\begin{equation*}
D\left( e^{ik\cdot x}\right) 
{{}^\circ}%
=ik\left( e^{ik\cdot x}\right) 
{{}^\circ}%
\end{equation*}%
might be violated in some point $x$ "at infinity" even if it holds for every 
$x\in \mathbb{R}^{N}$. In particular, by Axiom \ref{5}, it follows that $%
0\notin \sigma (\mathbf{p}).$

Next let us consider the generalization of the Laplacian operator%
\begin{equation*}
\Delta :C^{2}\left( \mathbb{R}^{N}\right) \rightarrow C^{0}\left( \mathbb{R}%
^{N}\right)
\end{equation*}%
defined by%
\begin{equation*}
D^{2}=\dsum\limits_{j=1}^{N}D_{j}^{2}:V%
{{}^\circ}%
\rightarrow V%
{{}^\circ}%
.
\end{equation*}

Probably the most important operator in classical QM is the following
Hamiltonian operator%
\begin{equation}
\mathbf{H}u(x)=-\frac{1}{2}\Delta u(x)+\mathbf{V}(x)u(x)\footnote{%
Here, for simplicity, we assume that $m_{i}=1$; $m_{i}$ is the mass of the $%
i $-th particle.}  \label{ham}
\end{equation}%
While in the $L^{2}$-theory a very delicate question is to choose an
appropriate potential $\mathbf{V}$ such that (\ref{ham}) make sense and to
define an appropriate selfadjoint realization of $\mathbf{H}$, in the theory
of ultrafunctions any internal function $\mathbf{V}:\Gamma \rightarrow 
\mathbb{E}$ provides a selfadjoint operator on $H%
{{}^\circ}%
$given by 
\begin{equation}
\mathbf{H%
{{}^\circ}%
}u(x):=-\frac{1}{2}D^{2}u(x)+\mathbf{V}(x)u(x)  \label{ham+}
\end{equation}%
with a spectrum consisting only of eigenvalues.

In particular it is possible to consider "very singular potential" such as%
\begin{eqnarray}
\mathbf{V}(x) &=&k\delta _{a}(x),\ \ k\in \mathbb{E}  \label{bibi} \\
\mathbf{V}(x) &=&\alpha \chi _{\Omega }(x)  \label{bobo}
\end{eqnarray}%
where $\alpha $, defined by (\ref{alfa}), is an infinite number. These
potentials cannot be defined outside of NAM; nevertheless they have an
interesting physical meaning.

\subsection{The heat equation\label{he}}

In QM and more in general in Mathematical Physics, an expression such as (%
\ref{ham}) does not define a selfadjoint operator; it is necessary to
construct a selfadjoint realization of it; sometimes it is a very delicate
question and it has relevant physical meaning. How can we compare this fact
with the NAM-theory where (\ref{ham+}) always describe a selfadjoint
operator with a spectrum of only eigenvalues? As matter of fact NAM-theory
can overcome these difficulties using a larger set of numbers (and hence of
operators) to describe the different physical phenomena. For example in (\ref%
{bobo}) the infinite number $\alpha $ is present.

In order to clarify this issue, let us consider a very simple example: the
heat equation in dimension 1%
\begin{equation*}
\partial _{t}u=\frac{1}{2}\partial _{x}^{2}u
\end{equation*}%
In this case we have that $\mathbf{H}=\frac{1}{2}\partial _{x}^{2}$ $\ $and $%
\mathbf{V}(x)=0.$ Suppose that we want do describe the diffusion of heat in
a bar modelled by the interval $\left[ 0,1\right] ;$ we suppose that the
initial condition is given by a function $\psi \in L^{2}\left( \left[ 0,1%
\right] \right) .$ This problem is not well posed because there are
different selfadjoint realizations of $\mathbf{H}$ in $L^{2}\left( \left[ 0,1%
\right] \right) $ which describe different different physical situations.
These selfadjoint realizations are determined by imposing boundary
conditions (BC); the most important ones are:

\begin{itemize}
\item the Dirichlet BC:%
\begin{equation*}
u(t,0)=u(t,1)=0
\end{equation*}

\item the Neumann BC:%
\begin{equation*}
\partial _{x}u(t,0)=\partial _{x}u(t,1)=0
\end{equation*}
\end{itemize}

These two kind of conditions provide two different selfadjoint realizations
of $\mathbf{H}$ which we will call $\mathbf{H}_{D}$ and $\mathbf{H}_{N}$
respectively. The spectrum of the first one is given by%
\begin{equation*}
\sigma \left( \mathbf{H}_{D}\right) =\pi \mathbb{N}=\left\{ \pi ,~2\pi
,~3\pi ,...\right\} 
\end{equation*}%
and the eigenfunctions are%
\begin{equation*}
\left\{ \sin \left( \pi x\right) ,~\sin \left( 2\pi x\right) ,~\sin \left(
3\pi x\right) ,...\right\} 
\end{equation*}%
The spectrum of the $\mathbf{H}_{N}$ is given by%
\begin{equation*}
\sigma \left( \mathbf{H}_{N}\right) =\pi \mathbb{N}_{0}=\left\{ 0,\ \pi
,~2\pi ,~3\pi ,...\right\} 
\end{equation*}%
and the eigenfunctions are%
\begin{equation*}
\left\{ 1,\ \cos \left( \pi x\right) ,~\cos \left( 2\pi x\right) ,~\cos
\left( 3\pi x\right) ,...\right\} 
\end{equation*}%
In the ultrafunction theory, the two physical phenomena are described by
choosing different Hamiltonian defined on all $V%
{{}^\circ}%
$:%
\begin{eqnarray}
\mathbf{H}_{D}^{\circ }u &=&-\frac{1}{2}D^{2}u+\alpha \chi _{\left[ 0,1%
\right] ^{c}}^{\circ }(x)u;\ \ \left[ 0,1\right] ^{c}=\mathbb{R}\backslash %
\left[ 0,1\right] \   \label{DD} \\
\mathbf{H}_{N}^{\circ }u &=&-\frac{1}{2}D\left( \chi _{\left[ 0,1\right]
}^{\circ }(x)Du\right)   \label{NN}
\end{eqnarray}%
The physical meaning of these two Hamiltonians is evident: in the first case
we have that, outside $\left[ 0,1\right] $, the heat is absorbed with an
infinite strength $-\alpha u$ so that it can reach only infinitesimal
values. In the second case we have that the diffusion coefficient $k(x)\chi
_{\left[ 0,1\right] }^{\circ }$ vanishes out of $\left[ 0,1\right] $ and
hence, all the heat is kept inside our bar. In both cases the evolution is
given by the exponential matrix%
\begin{equation*}
u(t,x)=e^{-t\mathbf{H}_{D}^{\circ }t}\psi 
{{}^\circ}%
;\ u(t,x)=e^{-t\mathbf{H}_{N}^{\circ }}\psi 
{{}^\circ}%
.
\end{equation*}

Concluding, in classical mathematics we have to choose different
self-adjoint realization of a given differential operator in order to
describe a given phenomenon. In ultrafunction theory, it is sufficient to
choose the appropriate operator since you have a larger set of operators.

\section{Ultrafunctions and Quantum Mechanics}

In this section we will describe an application of the previous theory to
the formalism of Quantum Mechanics. In the usual formalism, a physical state
is described by a unit vector $\psi \ $in a Hilbert space $\mathcal{H}$ and
an observable by a self-adjoint operator defined on it. In the
ultrafunctions formalism, a physical state is described by an ultrafunction $%
\psi \ $in $H%
{{}^\circ}%
=V%
{{}^\circ}%
+iV%
{{}^\circ}%
$ and an observable by a Hermitian operator defined on it.

The ultrafuctions approach to the QM-formalism presents the following
peculiarities:

\begin{itemize}
\item once you have learned the basic facts of $\Lambda $-theory, the
formalism which you get is easier to handle since it is based on Hermitian
matrices rather than on unbounded self-adjoint operators in Hilbert spaces;

\item this approach is closer to the "infinite" matrix approach of the
beginning of QM before the work of von Neumann and also closer to the way of
thinking of the theoretical physicists and chemists;

\item all observables (hyperfinite matrices) have infinitely many
eigenvectors; so the continuous spectrum can be considered as a set of
eigenvalues infinitely close to each other;

\item the operator%
\begin{equation*}
\mathbf{H%
{{}^\circ}%
}u(x)=-\frac{1}{2}D^{2}u(x)+\mathbf{V}(x)u(x)
\end{equation*}%
defines a selfadjoint operator on $H%
{{}^\circ}%
=V%
{{}^\circ}%
+iV%
{{}^\circ}%
$ for any ultrafunction $\mathbf{V}(x).$

\item the dynamics does not present any difficulty since it is given by the
exponential matrix relative to the Hamiltonian matrix $\mathbf{H%
{{}^\circ}%
}$;

\item the ideal ultrafunctions, namely the ultrafunctions which are not
close to any "classical state" such as the Dirac ultrafunctions, represent
ideal states which cannot be reproduced in laboratory.
\end{itemize}

\subsection{The axioms of Quantum Mechanics}

We start giving a list of the main axioms of quantum mechanics as it is
usually given in any textbook and then we will compare it with the
alternative formalism based on ultrafunctions.

\bigskip

{\large Von Neumann Axioms of QM}\bigskip

\textbf{Axiom C1}. A physical state is described by a unit vector $\psi \ $%
in a Hilbert space $\mathcal{H}.$

\bigskip

\textbf{Axiom C2.} An observable is represented by a self-adjoint operator $%
A $ on $\mathcal{H}$.

(a) The set of observable outcomes is given by the eigenvalues $\mu _{j}$ of 
$A$.

(b) After an observation/measurement of an outcome $\mu _{j}$, the system is
left in a eigenstate $\psi _{j}$ associated with the detected eigenvalue $%
\mu _{j}$.

(b) In a measurement the transition probability $\mathcal{P}$ from a state $%
\psi $ to an eigenstate $\psi _{j}$ is given by 
\begin{equation*}
\mathcal{P}=\left\vert \left( \psi ,\psi _{j}\right) \right\vert ^{2}.
\end{equation*}

\bigskip

\textbf{Axiom C3}. The evolution of a state is given by the Shroedinger
equation%
\begin{equation*}
i\frac{\partial \psi }{\partial t}=\mathbf{H}\psi
\end{equation*}%
where $H,\ $the Hamiltonian operator, is a self-adjoint operator
representing the energy of the system.

\bigskip

{\large Axioms of QM based on ultrafunctions}

\bigskip

\textbf{Axiom U1}. A physical state is described by a unit complex-valued
ultrafunction $\psi .$

\bigskip

\textbf{Axiom U2.} An observable is represented by a Hermitian operator $A$
on $H%
{{}^\circ}%
$.

(a) The set of observable outcomes is given by 
\begin{equation*}
st\left( \mu _{j}\right)
\end{equation*}
where $\mu _{j}$ is an eigenvalue of $A$.

(b) After an observation/measurement of an outcome $st\left( \mu _{j}\right) 
$, the system is left in an eigenstate $\psi _{j}$ associated with the
detected eigenvalue $\mu _{j}$.

(b) In a measurement the transition probability $\mathcal{P}$ from a state $%
\psi $ to an eigenstate $\psi _{j}$ is given by 
\begin{equation*}
\mathcal{P}=\left\vert \left( \psi ,\psi _{j}\right) \right\vert ^{2}.
\end{equation*}

\bigskip

\textbf{Axiom U3}. The evolution of the state of a system is given by the
Shroedinger equation%
\begin{equation}
i\frac{\partial \psi }{\partial t}=\mathbf{H%
{{}^\circ}%
}\psi  \label{sh+}
\end{equation}%
where $\mathbf{H%
{{}^\circ}%
},$ the Hamiltonian operator, representing the energy of the system.

\bigskip

\textbf{Axiom U4}. In laboratory you can realize only the states which
correspond to a finite expectation value of the energy (and/or the other
physically relevant quantities). We will call them physical states and the
others will be called ideal states.

\subsection{Discussion of the axioms}

AXIOM 1. In the classical formalism, a physical system is described by a
vector in an Hilbert space. In particular, taking the Schroedinger
representation of $\mathcal{H}$, $\psi $ can be represented by a function $%
\psi \in L^{2}(\Omega ),\ \Omega \subset \mathbb{R}^{N}$; so, by Def. \ref%
{L2} there is the following canonical embedding 
\begin{equation*}
{{}^\circ}%
:\mathcal{H}\rightarrow H%
{{}^\circ}%
\end{equation*}%
\begin{equation*}
\psi \mapsto \psi 
{{}^\circ}%
\end{equation*}%
Since $H%
{{}^\circ}%
$ is much richer than $\mathcal{H}$, in the ultrafunction framework there
exist more possible states; in particular the ideal states, see Axiom 
\textbf{U4.}

\bigskip

AXIOM 2. In the ultrafunction formalism, the Von Neuman notion of
self-adjoint operator is not needed. In fact observables can be represented
by internal Hermitian operators which are trivially self-adjoint. It follows
that any observable has exactly $\kappa =\dim ^{\ast }(H%
{{}^\circ}%
)$ eigenvalues (of course, if you take account of their multiplicity). No
essential distinction between eigenvalues and continuous spectrum is
required. For example, consider the eigenvalues of the position operator $%
\mathbf{q}$ of a free particle. The eigenfunction relative to an eigenvalue $%
q\in \mathbb{R}$ is the Dirac ultrafunction $\delta _{q}$

The eigenvalues of an internal Hermitian operator $A$ are Euclidean numbers,
and hence, assuming that a measurement gives a real number, we have imposed
in Axiom 2 that the outcome of an experiment is $st(\mu )$. However, we
think that the probability is better described by the Euclidean number $%
\left\vert \left( \psi ,\psi _{j}\right) \right\vert ^{2}$ rather than the
real number $st(\left\vert \left( \psi ,\psi _{j}\right) \right\vert ^{2})$
. For example, let $\psi \in H%
{{}^\circ}%
$ be the state of a system; the probability of that a particle is in the
position $q$ is given by%
\begin{equation*}
\left\vert \doint \psi (x)\delta _{q}(x)\sqrt{d(a)}dx\right\vert =\left\vert
\psi (q)\right\vert \sqrt{d(q)}
\end{equation*}%
which is is an infinitesimal number. We refer to \cite{BHW,BHW16} for a
presentation and discussion of the Non Archimedean Probability.

\bigskip

AXIOM 3. Since $\mathbf{H%
{{}^\circ}%
}$ is an internal operator defined on a hyperfinite vector space, it can be
represented by an Hermitian hyperfinite matrix and hence the evolution
operator of (\ref{sh+}) is described by the exponential matrix $e^{it\mathbf{%
H%
{{}^\circ}%
}}.$

\bigskip

AXIOM 4. In ultrafunction theory, the mathematical distinction between
physical eigenstates and the ideal eigenstates is intrinsic and it does not
correspond to anything in the usual formalism. The point is to know if it
corresponds to something physically meaningful. Basically, we can say that
the physical states can be prepared and measured in a laboratory, while the
ideal states represent "extreme" situations useful in the foundations of the
theory, in thought experiments (gedankenexperiment) and in the computations.
For example the Dirac ultrafunction is not a physical state but an ideal
state and it represents a situation in which the position of a particle is
perfectly determined. Clearly this state cannot be produced in a laboratory
since it requires infinite energy (see section \ref{ha}), but nevertheless
it is useful in our description of the physical world. This situation makes
more explicit something which is already present in the classical approach.
For example, in the Shroedinger representation of a free particle in $%
\mathbb{R}^{3}$, consider the state 
\begin{equation*}
\psi (x)=\frac{\varphi (x)}{|x|},\ \varphi \in C(\mathbb{R}^{3}),\ \varphi
(0)>0.
\end{equation*}%
We have that $\psi \in L^{2}(\mathbb{R}^{3})$ but this state cannot be
produced in a laboratory, since the expected value of its energy 
\begin{equation*}
\left\langle \mathbf{H}\psi ,\psi \right\rangle =\frac{1}{2}\int \left\vert
\nabla \psi \right\vert ^{2}dx
\end{equation*}%
is infinite (even if the result of a single experiment is a finite number).
In other words, Axiom $\mathbf{U4}$ makes formally precise something which,
in some sense, is already present (but hidden) in the classical formalism.

\bigskip

\subsection{The Heisenberg algebra\label{ha}}

\bigskip

In this section, we will apply ultrafunction theory to the description of a
quantum particle via the algebraic approach. For simplicity here we consider
the one-dimensional case. The states of a particle are defined by the
observables $q$ and $p$ which represent the position and the momentum
respectively. A quantum particle is described by the algebra of observables
generated by $p$ and $q$ according to the following commutation rules:%
\begin{equation*}
\left[ p,q\right] =i,\ \ \left[ p,p\right] =0,\ \ \left[ q,q\right] =0
\end{equation*}%
The algebra generated by $p$ and $q$ with the above relations is called the
Heisenberg algebra and denoted by $\mathfrak{A}_{H}$. The Heisenberg algebra
does not fit in the general theory of $C^{\ast }$-algebras since both $p$
and $q$ are not bounded operator. The usual technical solution to this
problem is done via the Weyl operators and the Weyl algebra (for more
details and a discussion on this point we refer to \cite{strocchi05}).

Let us see an alternative approach via the ultrafunction theory. First of
all we take the following representation of $\mathfrak{A}_{H}\ $%
\begin{equation*}
J:\mathfrak{A}_{H}\rightarrow \mathfrak{A}(H%
{{}^\circ}%
)
\end{equation*}%
where $\mathfrak{A}(H%
{{}^\circ}%
)$ is the algebra of the Hermitian internal operators on $H%
{{}^\circ}%
$; $J$ is defined by%
\begin{equation*}
J(p)=\mathbf{p}=-iD;\ \ J(q)=\mathbf{q}=x.
\end{equation*}%
(see (\ref{pi}) and (\ref{qu})). $\mathbf{p}$ and $\mathbf{q}$ are Hermitian
operators and hence $H%
{{}^\circ}%
$ has an orthonormal basis generated by the eigenfunctions of $\mathbf{p}$
or $\mathbf{q}$. A very interesting fact is that the ultrafunction model
violate the Heisenberg uncertainty relations $\left[ \mathbf{p},\mathbf{q}%
\right] =i$. To see this fact, we argue indirectly. Assume that the
Heisenberg relation holds; then 
\begin{equation*}
\left\langle \left[ \mathbf{p},\mathbf{q}\right] \delta _{a},\delta
_{a}\right\rangle =i\left\Vert \delta _{a}\right\Vert ^{2}.
\end{equation*}%
On the other hand, by a direct computation\footnote{%
Notice that this computation implies the violation of the Leibniz rule of
the differentation of a product. This fact is consistent with the Schwartz
impossibility theorem \cite{Schwartz} which states that there is not a
differential algebra containing a set isomorphic to the space of
distribution. In our case, we have that $V%
{{}^\circ}%
$ is an algebra "containing" the distributions and hence, it cannot be a
differential algebra; in fact the Leibniz rule, in some cases, is violated. }%
,\label{leib} we get:%
\begin{eqnarray*}
\doint \left[ \mathbf{p},\mathbf{q}\right] \delta _{a}(x)\overline{\delta
_{a}(x)}~dx &=&\doint -iD\left( x\delta _{a}(x)\right) \delta
_{a}(x)~dx+\doint x\left( iD\delta _{a}(x)\right) \delta _{a}(x)~dx \\
&=&-ia\doint \left[ D\delta _{a}(x)\right] \delta _{a}(x)~dx+i\doint x\left[
D\delta _{a}(x)\right] \delta _{a}(x)~dx \\
&=&-iaD\delta _{a}(a)+iaD\delta _{a}(a)=0
\end{eqnarray*}

This fact is consistent with the Axiom \textbf{U4 }which establishes that
the ideal states cannot be produced in laboratory. According to this
description of QM, the uncertainty relations hold only for the limitation of
the experimental apparatus. In a laboratory you can prepare a state
corresponding to a function 
\begin{equation*}
\psi (x)=\sum_{a\in \Gamma }\psi \left( a\right) \chi _{a}(x)=\sum_{a\in
\Gamma }\frac{\psi \left( a\right) }{d(a)}\delta _{a}(x)
\end{equation*}%
for which the expectation value of the energy is finite. But the eigenfuncion%
\begin{equation*}
\psi _{a}(x)=\frac{\delta _{a}(x)}{\left\Vert \delta _{a}\right\Vert }=\sqrt{%
d(a)}\delta _{a}(x)
\end{equation*}%
has an infinite expectation value of the energy as the following computation
shows: 
\begin{eqnarray*}
\left\langle \mathbf{H%
{{}^\circ}%
}\psi ,\psi \right\rangle &=&\frac{1}{2}\int \left\vert D\psi
_{a}\right\vert ^{2}dx=\frac{1}{2}d(a)\int \left\vert D\delta
_{a}\right\vert ^{2}dx \\
&=&\frac{1}{2}d(a)\left\Vert D\delta _{a}\right\Vert ^{2}=\frac{\left\Vert
D\delta _{a}\right\Vert ^{2}}{\left\Vert \delta _{a}\right\Vert ^{2}}
\end{eqnarray*}%
The conclusion follows from the following proposition:

\begin{proposition}
$\frac{\left\Vert D\delta _{q}\right\Vert }{\left\Vert \delta
_{q}\right\Vert }$ is an infinite number.
\end{proposition}

\textbf{Idea of the proof}: The Poincar\'{e} inequality states that%
\begin{equation*}
\forall u\in C_{c}^{1}(a,b),\ \ \int_{a}^{b}\left\vert u\right\vert
^{2}dx\leq \left( b-a\right) ^{2}\int_{a}^{b}\left\vert \partial
u\right\vert ^{2}dx
\end{equation*}%
Since $\delta _{a}$ is the $\Lambda $-limit of functions having support in
an interval whose length tends to $0$, then 
\begin{equation*}
\int_{a}^{b}\left\vert \delta _{q}\right\vert ^{2}dx\leq \varepsilon
^{2}\int_{a}^{b}\left\vert D\delta _{q}\right\vert ^{2}dx
\end{equation*}%
where $\varepsilon $ is an infinitesimal.

$\square $

In conclusion we get the following picture. The ultrafunction Heisenberg
algebra $\mathfrak{A}(H%
{{}^\circ}%
)$ is an algebra over the field $\mathbb{C}^{\ast }$ and hence it is also an
algebra over $\mathbb{C}$ and $\mathfrak{A}_{H}$ is a subalgebra of $%
\mathfrak{A}(H%
{{}^\circ}%
)$ over the field $\mathbb{C}$. One of the advantages in using $\mathfrak{A}%
(H%
{{}^\circ}%
)$ is that all the operators in $\mathfrak{A}(H%
{{}^\circ}%
)$ are bounded and the spectrum consists of eigenvalues, namely it behaves
as a matrix algebra. The price that we pay for this is the existence of
ideal states which are not is not feasible in a laboratory. With this
respect, the Heisenberg uncertainty relations take a somewhat different
philosophical meaning.

\section{Conclusion}

If we live in a larger space there are more things to see and more things we
can do. This is the palpable truth that we tried to illustrate in this work.

If we work in a mathematical universe containing infinitesimal, we can have
a greater choice for models able to describe natural phenomena. In
particular, we can describe the physical space as a set of points wider than
those that can be described by real coordinates. Furthermore we can make
space models whose points are represented by a hyperfinite grid. In this way
we can have a set of functions much wider than the real functions, but at
the same time they are easier to handle because they satisfy the same
(internal) rules of the vectors in a finite dimensional vector space. The
use of this space model has many advantages in a wide range of problems. In
particular, the axioms of the QM that we have illustrated here allow
considerable technical simplifications by reducing the self-adjoint
operators to Hermitian matrices of hyperfinite size. Among the possible
eigenstates of an unbounded operator, there are also those who have infinite
energy, which although not achievable in a laboratory allow a broader vision
and an alternative description of the phenomenon. For example, Heisemberg's
relations are reformulated as follows: \textit{in order to determine exactly
the position of a particle, an experimental apparatus is needed that can
provide the particle with an infinite impulse and energy.}

In conclusion, we belive that NAM in the future can have more uses than
commonly think, both in foundational and applicative.

\bigskip

\bigskip

\textsc{V. Benci - Dipartimento di Matematica}

\textsc{Universita degli Studi di Pisa,}

\textsc{Via F. Buonarroti 1/c, 56127 Pisa, Italy}

\textsc{and Centro Linceo interdisciplinare Beniamino Segre}

\textsc{Palazzo Corsini - Via della Lungara 10, 00165 Roma, Italy}

\textsc{E-mail address: benci@dma.unipi.it}

\end{document}